%% file: manuscript.tex
\documentclass[letter,twoside,web]{ieeecolor}
\pdfoutput=1
\input{setup}

\begin{document}
\bstctlcite{IEEEexample:BSTcontrol}
\title{Two-Temperature Principle for Electrothermal Performance Evaluation of GaN HEMTs}
\author{Yang Shen, and Bing-Yang Cao
	\thanks{This work was supported by the National Natural Science Foundation of China (Nos. 51825601, U20A20301, 52250273). (Corresponding author: Bing-Yang Cao.)}
	\thanks{Yang Shen and Bing-Yang Cao are with the Key Laboratory of Thermal Science and Power Engineering of Education of Ministry, Department of Engineering Mechanics,
	Tsinghua University, Beijing 100084, China (e-mail: sheny21@mails.tsinghua.edu.cn; caoby@tsinghua.edu.cn).}}
\markboth{Shen \MakeLowercase{\textit{et al.}}: Two-Temperature Principle for Electrothermal Performance Evaluation of GaN HEMTs}%
{Shen \MakeLowercase{\textit{et al.}}: Two-Temperature Principle for Electrothermal Performance Evaluation of GaN HEMTs}


\maketitle

\begin{abstract}
	We present a comprehensive investigation of self-heating in gallium nitride (GaN) high-electron-mobility transistors (HEMTs) through technology computer-aided design (TCAD) simulations and phonon Monte Carlo (MC) simulations.
	With microscopic phonon-based electrothermal simulations, we scrutinize both the temperature profiles and electrothermal coupling effect within GaN HEMTs.
	Two metrics, maximum channel temperature ($T_\text{max}$) and equivalent channel temperature ($T_\text{eq}$), are introduced to measure the reliability and electrical performance degradation of the device, respectively.
	The influence of bias-dependent heat generation and phonon ballistic transport on the two indicators is thoroughly examined.
\end{abstract}


\section{Introduction}

	Gallium Nitride (GaN) high electron mobility transistors (HEMTs) have emerged as exceptional devices for high-frequency and high-power applications \cite{meneghini2021gan, haziq2022challenges}.
	However, these devices encounter a significant thermal bottleneck due to high power dissipation, leading to severe reliability issues and electrical performance degradation \cite{rosker2009darpa, padmanabhan2013reliability,ranjan2019investigation}.
	As external cooling techniques have matured over time, the focus has shifted towards near-junction thermal management strategies, \eg{}, directly modifying the device designs to mitigate the overheating \cite{cho2015near,choi2021perspective,chatterjee2021electro}.

	Nonetheless, a significant limitation exists in the current landscape of device thermal analysis and optimization, where the prevalent approach only employs the maximum channel temperature ($T_\text{max}$) as the benchmark for device thermal resistance, with a primary focus on minimizing $T_\text{max}$ \cite{kim2023modeling,song2020effect}.
	However, Chen \etal{} discovered that $T_\text{max}$ does not directly correlate with the electrical performance of GaN HEMTs \cite{chen2019self}.
	Instead, the equivalent channel temperature ($T_\text{eq}$), which is close to the average channel temperature, is a more appropriate metric for assessing self-heating-induced electrical performance degradation. 
	Therefore, it is crucial to accurately evaluate both $T_\text{max}$ and $T_\text{eq}$ for various device designs since they can significantly differ even for the same device.
	However, the current thermal simulation practices have two major shortcomings that may affect the evaluation of these metrics. 
	First, the heat source is typically modeled as a surface heat flux or a volumetric heat generation at the top of the GaN buffer layer with a fixed length \cite{Gerrer2021, Bagnall2014}, where the bias dependence is often overlooked \cite{chen2020modeling,odabacsi2020improved}.
	Also, the majority of existing works rely on Fourier's law-based finite element method (FEM) \cite{hua2019thermal,shen2022bias,chatterjee2020nanoscale}.
	Whereas, in GaN HEMTs the characteristic size is comparable to the mean free paths (MFP) of phonons, which are primary heat carriers in semiconductors \cite{chen2021non,bao2018review,tang2023phonon}.
	In this case, Fourier's law becomes inapplicable, resulting in potentially unreliable thermal designs \cite{hao2018electrothermal, SHEN2023124284,vermeersch2022thermal}.
	These compounding issues have resulted in varying outcomes across different studies, introducing ambiguity into thermal design rules.

	Therefore, it is essential to conduct a comprehensive examination of both the thermal metrics in GaN HEMTs with all these factors incorporated, giving broader insights and guidelines for device thermal designs.
	In this study, we provide a thorough discussion of the issues through microscopic electrothermal simulations of GaN HEMTs by integrating technology computer-aided design (TCAD) simulations and phonon Monte Carlo (MC) simulations.

\section{Device Structure and Simulation Setup}

	\begin{figure}[htbp!]
		\centering
		\subfloat{\includegraphics[width=0.48\linewidth]{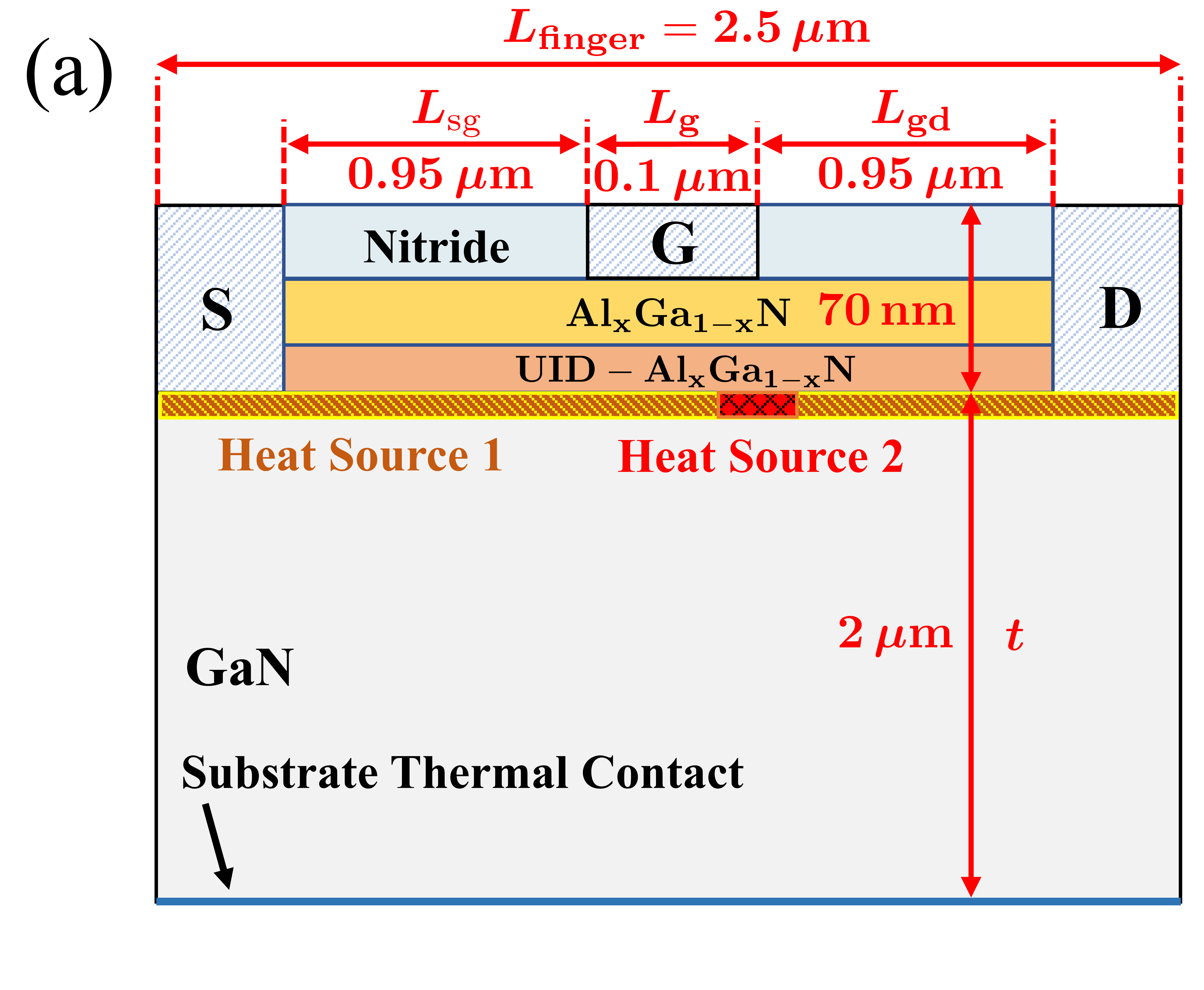}}
		\subfloat{\includegraphics[width=0.49\linewidth]{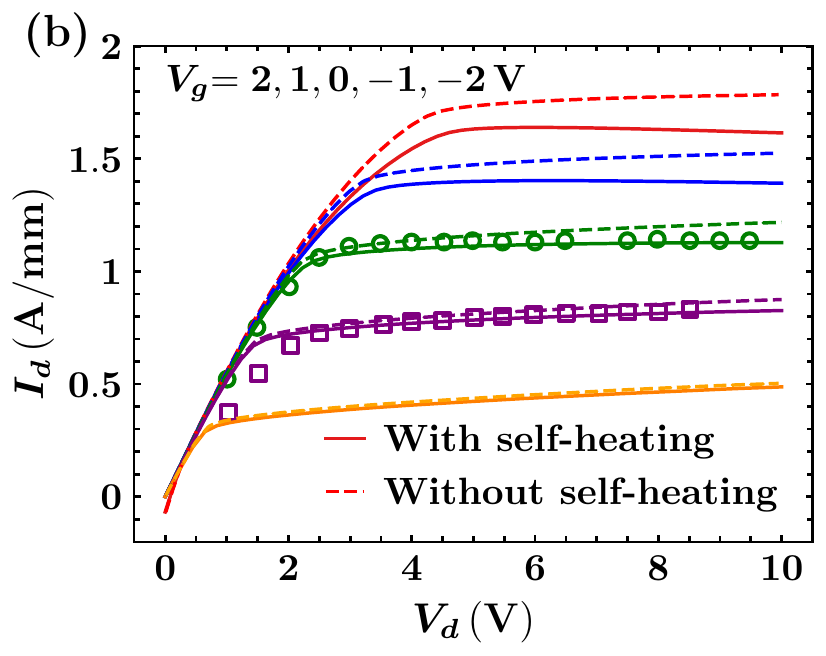}}
		\caption{(a) Schematic of the GaN HEMT. (b) Output characteristics of the HEMT extracted from TCAD simulations with film thermal conductivity (lines) and experimental results (symbols).}
		\label{fig:schematic}
	\end{figure}

	The AlGaN/GaN HEMT structure under examination is depicted in Fig. \ref{fig:schematic} (a), which replicates the device presented in Ref. \cite{jatal2014high}.
	Sentaurus TCAD is used for device construction and simulation.
	The Masetti model \cite{masetti1983modeling} is adopted to account for doping dependence of electron mobility, and the Canali model \cite{canali1975electron} is used to cover temperature and electric field dependence.
	Detailed setup information is available in Ref. \cite{chen2019self,chen2020modeling}.
	Figure \ref{fig:schematic} (b) illustrates the simulated output characteristics of the HEMT.
	Accounting for self-heating effects, the drain current $I_d$ degrades, aligning well with the experimental DC results.
	With the heat generation profiles predicted by TCAD simulations as phonon emission sources, phonon MC simulations can be conducted to solve the phonon Boltzmann Transport Equation (BTE)  \cite{hao2017hybrid,warzoha2021applications}.
	Further details regarding the simulation process can be found in Ref. \cite{shen2022bias}.

\section{Results and Dissusion}

\subsection{$\textit{T}_\text{max}$ and $\textit{T}_\text{eq}$ Analysis}

\begin{figure}[htbp!]
		\centering
		\includegraphics[width=0.8\linewidth]{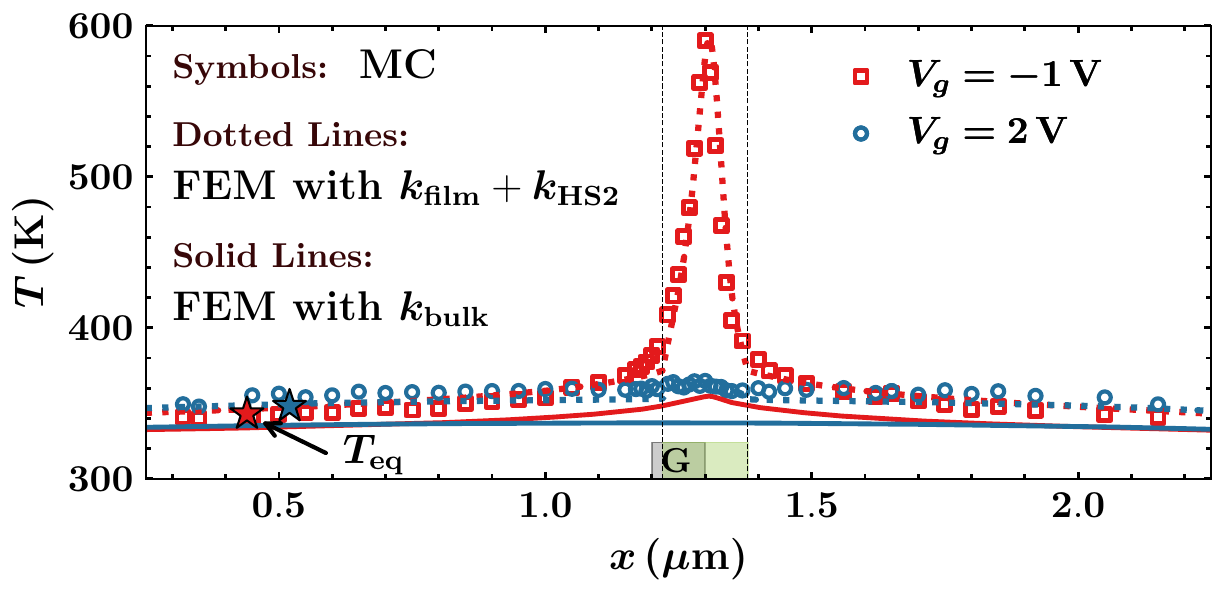}
		\caption{Channel temperature profiles predicted by MC simulations (symbols), and FEM with $k_\text{bulk}$ and $k_\text{film} + k_\text{HS2}$, respectively, at different biases with $P_\text{diss} = \SI{5}{\watt/\mm}$. The biasing points are $(V_g, V_d) = (\SI{-1}{\volt}, \SI{6.7}{\volt}), (\SI{1}{\volt}, \SI{4.1}{\volt})$, and $(\SI{2}{\volt}, \SI{3.8}{\volt})$. The positions of the gate and the high-field region are marked in the figure.}
		\label{fig:original_temperature}
	\end{figure}

	Figure \ref{fig:original_temperature} depicts the channel temperature profiles as predicted by MC simulations and FEM across different biases, at an identical power dissipation of $P_\text{diss} = \SI{5}{\watt/\mm}$.
	Across the channel, MC simulations consistently yield higher temperatures than FEM with bulk thermal conductivity ($k_\text{bulk}$) due to the influence of phonon ballistic transport.
	At the drain-side gate edge, $T_\text{max}$ predicted by MC are markedly higher and demonstrate significant bias-dependence.
	Conversely, the temperature differences in the drain and source access regions are less dramatic, and bias dependence is less evident.
	These diverse behaviors stem from disparate phonon ballistic transport mechanisms.
	The cross-plane ballistic effect, triggered by phonon-boundary scattering, depends mainly on the GaN layer's thickness and can uniformly elevate channel temperatures \cite{hua2019thermal}.
	In contrast, the ballistic effect when the heat source size is comparable to phonon MFP, heavily relies on heat source size and primarily increases the heat source temperature \cite{hao2018electrothermal}.
	Chen \etal{} observed that in the linear regime, when the drain-to-source voltage $V_d$ is less than the saturation voltage $V_\text{dsat}$, heat is uniformly dissipated in the whole finger, denoted as Heat Source 1 (HS1), as shown in Fig. \ref{fig:schematic} (a) \cite{chen2020modeling}.
	When $V_d > V_\text{dsat}$, heat dissipation in HS1 remains at its peak, and new heat generation occurs exclusively at the drain-side gate edge, denoted as Heat Source 2 (HS2) with a much narrower region of $L_\text{HS2} \approx \SI{160}{\nm}$.
	The heat source-related ballistic effect becomes significant only when the heat begins to dissipate in HS2 with a much smaller size.

	Hence, to accommodate both types of phonon ballistic effects in TCAD simulations, two effective thermal conductivities are introduced to replace the $k_\text{bulk}$.
	The cross-plane effective thermal conductivity $k_\text{film}$ of \SI{120}{\watt/\m \kelvin} is adopted to account for the cross-plane ballistic effect.
	Additionally, an extremely low thermal conductivity $k_\text{HS2}$ of \SI{8}{\watt/\m \kelvin} is set in the HS2 region to account for the ballistic effect with the heat source size comparable to MFP.
	As shown in Fig. \ref{fig:original_temperature}, the channel temperature profiles predicted FEM with $k_\text{film}$ and $k_\text{HS2}$ show excellent concurrence with MC simulations across different biases.
	By reconstructing the temperature profiles in TCAD simulations, the influence of temperatures on the device's electrical performance can be analyzed.
	$T_\text{eq}$ can be used to correlate channel temperatures with device drain current degradation \cite{chen2019self}.
	It is defined as the uniform temperature ($T_\text{uniform}$) in which a device is immersed, such that its drain current matches the simulated drain current at the same bias when considering self-heating effects,
	\begin{equation}
		T_{\mathrm{eq}}\left(V_{\mathrm{GS}}, V_{\mathrm{DS}}\right)=\left.T_{\text{uniform}}\left(V_{\mathrm{GS}}, V_{\mathrm{DS}}\right)\right|_{@ I_{\mathrm{DS},\text{self-heating}}=I_{\mathrm{DS},\text{uniform}}}.
	\end{equation}
	Figure \ref{fig:original_temperature} marks the locations where the MC-predicted temperature equals $T_\text{eq}$ at different biases.
	The values and positions of $T_\text{eq}$ in different cases are almost identical, all lying far from the hotspot region.
	It implies that thorough the phonon ballistic transport can dramatically increase $T_\text{max}$, its impact on the device's electrical performance is not such significant.

	\begin{figure}[htbp!]
		\centering
		\subfloat{\includegraphics[width=0.5\linewidth]{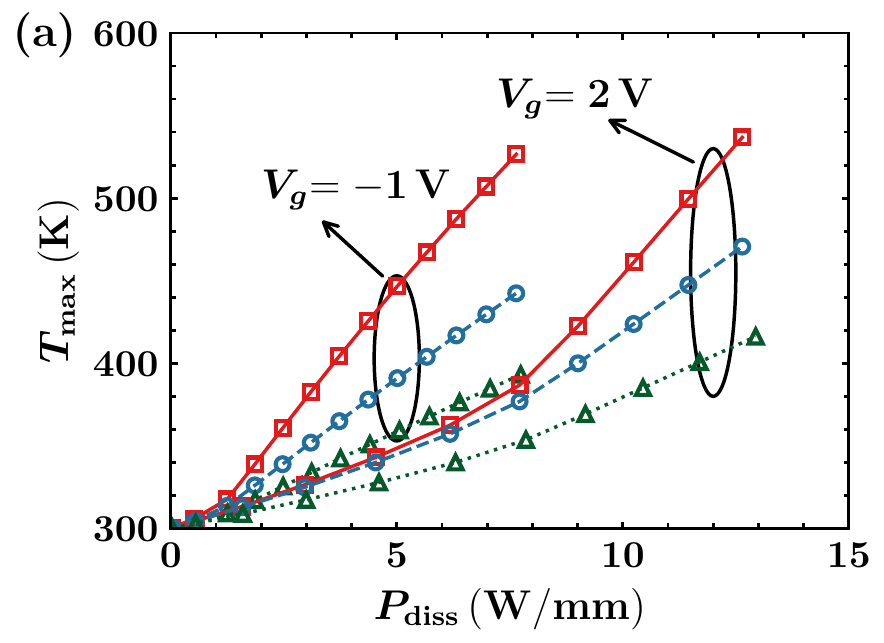}}
		\subfloat{\includegraphics[width=0.5\linewidth]{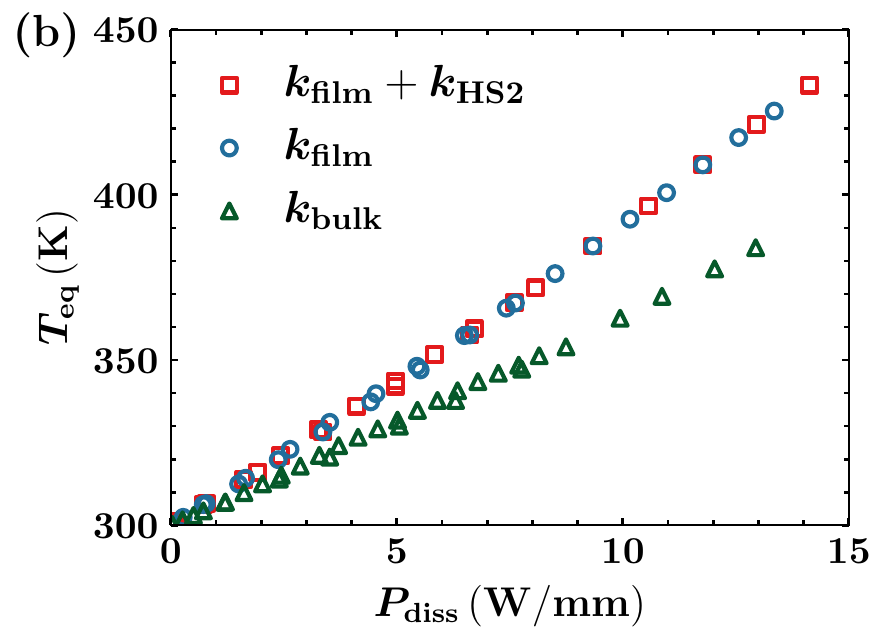}}
		\caption{(a) $T_\text{max}$ and (b) $T_\text{eq}$ varying with $P_\text{diss}$ under two biases for various thermal conductivity settings.} 
		\label{fig:short_channel_Teq}
	\end{figure}

	To thoroughly investigate the correlation between bias dependence, phonon ballistic effects, and the two temperature metrics, three parallel TCAD simulations are executed—namely, simulations with $k_\text{bulk}$, $k_\text{film}$, and $k_\text{film} + k_\text{HS2}$.
	Figure \ref{fig:short_channel_Teq} delineates the variation of $T_\text{max}$ and $T_\text{eq}$ to $P_\text{diss}$ with $V_g = \SI{-1}{\volt}$ and $\SI{2}{\volt}$, respectively, for various thermal conductivity settings.
	As demonstrated in Fig. \ref{fig:short_channel_Teq} (a), $T_\text{max}$ manifests a strong bias dependence, with every curve being distinctly bifurcated into linear and saturation regimes at $V_\text{dsat}$.
	Notably, at $P_\text{diss} = \SI{7.5}{\watt/\mm}$, even for the case utilizing $k_\text{bulk}$ , $T_\text{max}$ difference for $V_g = \SI{-1}{\volt}$ and $V_g = \SI{2}{\volt}$  reaches nearly \SI{50}{\kelvin}.
	Incorporating the cross-plane ballistic effect by substituting $k_\text{bulk}$ with $k_\text{film}$ results in a consistent increase in $T_\text{max}$ across both the linear and saturation regimes.
	The further integration of $k_\text{HS2}$ introduces a minor difference in the linear regime when compared with using only $k_\text{film}$.
	However, once $V_d > V_\text{dsat}$, $T_\text{max}$ and its bias dependence rise significantly due to the heat source-related ballistic effect, increasing the difference substantially to around \SI{100}{\kelvin}.
	In contrast, Fig. \ref{fig:short_channel_Teq} (b) reveals that $T_\text{eq}$ is substantially lower than $T_\text{max}$ at equivalent $P_\text{diss}$ and exhibits almost no bias dependence.
	The $T_\text{eq}$ results for the cases with $k_\text{bulk}$ and $k_\text{bulk} + k_\text{HS2}$ align closely and are higher than when using $k_\text{bulk}$ alone.

\subsection{Electrothermal Co-Analysis and Design Guidelines}

\begin{figure}[htbp!]
		\centering
		\includegraphics[width=\linewidth]{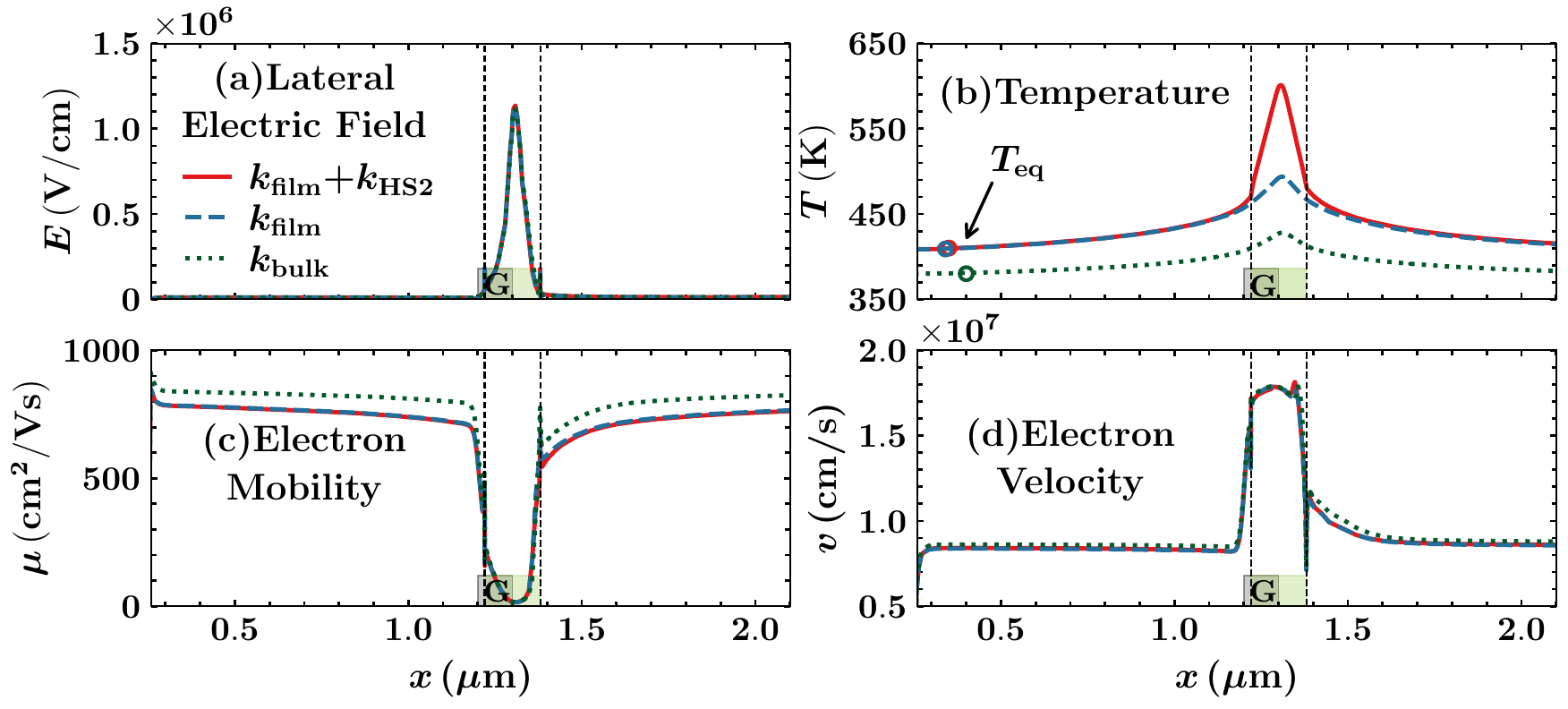}
		\caption{Distributions of (a) lateral electric field, (b) temperature, (c) electron mobility, and (d) electron velocity along the HEMT channel with different thermal conductivity settings at $V_g = \SI{0}{\volt}, V_d = \SI{10}{\volt}$.}
		\label{fig:short_channel_profile}
	\end{figure}

	To clarify the origin of different behaviors of $T_\text{max}$ and $T_\text{eq}$, Fig. \ref{fig:short_channel_profile} (a) - (d) show the TCAD-predicted distributions of the channel's lateral electric field, temperature, electron mobility, and electron velocity for different thermal conductivity settings.
	Electric field distributions remain nearly identical across all instances, however,	Utilizing $k_\text{film}$—commonly used in device thermal simulations—causes a uniform increase in the whole channel temperature compared to the case with $k_\text{bulk}$.
	The electron mobility and velocity in the access regions are degraded due to elevated temperatures.
	In the case with $k_\text{film} + k_\text{HS2}$, the heat source-related ballistic effect only raises the temperature in the HS2 region.
	The temperature in the rest of the channel closely matches the case with $k_\text{film}$.
	Despite the temperature elevation in the HS2 region, the electron mobility and velocity remain almost unchanged.
	The phenomenon can be attributed to the fact that in the low-field access region, electron mobility is governed by phonon scattering, significantly degrading as the temperature rises \cite{chen2019self}.
	Whereas in the high-field region, the electric field is strong enough to make the drift velocity reach its saturation value, which is weakly dependent on temperature.

	From the above analysis, the interplay among the two thermal metrics $T_\text{max}$ and $T_\text{eq}$, bias dependence of heat generation, and phonon ballistic transport becomes clear.
	$T_\text{max}$, linked to reliability concerns, is highly bias-dependent and is influenced by both the heat source-related ballistic effect and the cross-plane effect.
	Conversely, $T_\text{eq}$, which pertains to device drain current degradation, is nearly bias-insensitive and is exclusively affected by the cross-plane ballistic effect.
	To improve device reliability and abate self-heating-induced electrical performance degradation, it is indeed to enhance device designs to mitigate $T_\text{max}$ and $T_\text{eq}$, respectively.
	In device thermal optimizations, The size of the heat source can decisively influence the optimization of thermal designs in GaN HEMTs.
	As the size of heat source decreases, the thermal spreading resistance plays a more deterministic role than one-dimensional thermal resistance, dominating the near-junction heat transfer, enhancing the interfacial thermal resistance, and changing the optimization trends \cite{cho2015near}.
	Given substantial disparities in the equivalent widths for $T_\text{max}$ and $T_\text{eq}$, it is crucial to simultaneously contemplate these distinctions to identify an optimal design strategy.

\section{Conclusion}

	This letter presents a thorough investigation into self-heating effects in GaN HEMTs by integrating TCAD simulations and phonon MC simulations.
	It reveals that $T_\text{max}$ is highly bias-dependent, and is influenced by both the cross-plane effect and the heat-source-related ballistic effect, especially in the saturation regime.
	In contrast, $T_{eq}$ displays almost no bias dependence and is solely influenced by the cross-plane ballistic effect.
	To improve the device's electrothermal performance, it is essential to find a device design that can simultaneously mitigate both $T_\text{max}$ and $T_\text{eq}$.
	The study provides an in-depth understanding of the self-heating phenomenon in GaN HEMTs and may offer practical insights for device designs.
	

\bibliographystyle{IEEEtran}
\bibliography{reference.bib}

\end{document}

%% file: setup.tex
\usepackage{xspace}
\usepackage[x11names, dvipsnames, svgnames]{xcolor}
\usepackage[caption=false,font=normalsize,labelfont=sf,textfont=sf]{subfig}
\usepackage[separate-uncertainty=true,multi-part-units=single]{siunitx}
\usepackage{amsmath,amsfonts}
\usepackage{adjustbox} 
\usepackage{algorithmic}
\usepackage{algorithm}
\usepackage{array}
\usepackage{booktabs}
\usepackage[defaultcolor=red, final]{changes}
\usepackage{cite}
\usepackage{generic}
\usepackage{graphicx}
\usepackage{mathrsfs}
\usepackage{multirow}
\usepackage{mhchem}
\usepackage{stfloats}
\usepackage{textcomp}
\usepackage{url}
\usepackage{verbatim}
\usepackage{etoolbox}
\makeatletter
\@ifundefined{color@begingroup}%
  {\let\color@begingroup\relax
   \let\color@endgroup\relax}{}%
\def\fix@ieeecolor@hbox#1{%
  \hbox{\color@begingroup#1\color@endgroup}}
\patchcmd\@makecaption{\hbox}{\fix@ieeecolor@hbox}{}{\FAILED}
\patchcmd\@makecaption{\hbox}{\fix@ieeecolor@hbox}{}{\FAILED}
\makeatletter
\DeclareRobustCommand\onedot{\futurelet\@let@token\@onedot}
\def\@onedot{\ifx\@let@token.\else.\null\fi\xspace}

\def\eg{\emph{e.g}\onedot}

\def\etal{\emph{et al}\onedot}
\makeatother

\def\BibTeX{{\rm B\kern-.05em{\sc i\kern-.025em b}\kern-.08em
    T\kern-.1667em\lower.7ex\hbox{E}\kern-.125emX}}
\definechangesauthor[name={R1-C1}, color=blue]{R1-C1}
\definechangesauthor[name={R1-C2}, color=blue]{R1-C2}
\definechangesauthor[name={R1-C3}, color=blue]{R1-C3}
\definechangesauthor[name={R1-C4}, color=blue]{R1-C4}
\definechangesauthor[name={R2-C1}, color=red]{R2-C1}
\definechangesauthor[name={R2-C2}, color=red]{R2-C2}
\definechangesauthor[name={R2-C3}, color=red]{R2-C3}
\definechangesauthor[name={R2-C4}, color=red]{R2-C4}
\definechangesauthor[name={R2-C5}, color=red]{R2-C5}

%% file: manuscript.bbl
\begin{thebibliography}{10}
\providecommand{\url}[1]{#1}
\csname url@samestyle\endcsname
\providecommand{\newblock}{\relax}
\providecommand{\bibinfo}[2]{#2}
\providecommand{\BIBentrySTDinterwordspacing}{\spaceskip=0pt\relax}
\providecommand{\BIBentryALTinterwordstretchfactor}{4}
\providecommand{\BIBentryALTinterwordspacing}{\spaceskip=\fontdimen2\font plus
\BIBentryALTinterwordstretchfactor\fontdimen3\font minus
  \fontdimen4\font\relax}
\providecommand{\BIBforeignlanguage}[2]{{%
\expandafter\ifx\csname l@#1\endcsname\relax
\typeout{** WARNING: IEEEtran.bst: No hyphenation pattern has been}%
\typeout{** loaded for the language `#1'. Using the pattern for}%
\typeout{** the default language instead.}%
\else
\language=\csname l@#1\endcsname
\fi
#2}}
\providecommand{\BIBdecl}{\relax}
\BIBdecl

\bibitem{meneghini2021gan}
M.~Meneghini, C.~De~Santi, I.~Abid, M.~Buffolo, M.~Cioni, R.~A. Khadar,
  L.~Nela, N.~Zagni, A.~Chini, F.~Medjdoub \emph{et~al.}, ``{GaN}-based power
  devices: Physics, reliability, and perspectives,'' \emph{Journal of Applied
  Physics}, vol. 130, no.~18, p. 181101, 2021.

\bibitem{haziq2022challenges}
M.~Haziq, S.~Falina, A.~A. Manaf, H.~Kawarada, and M.~Syamsul, ``Challenges and
  opportunities for high-power and high-frequency {AlGaN/GaN}
  high-electron-mobility transistor ({HEMT}) applications: A review,''
  \emph{Micromachines}, vol.~13, no.~12, p. 2133, 2022.

\bibitem{rosker2009darpa}
M.~Rosker, C.~Bozada, H.~Dietrich, A.~Hung, D.~Via, S.~Binari, E.~Vivierios,
  E.~Cohen, and J.~Hodiak, ``The {DARPA} wide band gap semiconductors for {RF}
  applications ({WBGS-RF}) program: {Phase II} results,'' \emph{CS ManTech},
  vol.~1, pp. 1--4, 2009.

\bibitem{padmanabhan2013reliability}
B.~Padmanabhan, D.~Vasileska, and S.~Goodnick, ``Reliability concerns due to
  self-heating effects in {GaN HEMTs},'' \emph{Journal of Integrated Circuits
  and Systems}, vol.~8, no.~2, pp. 78--82, 2013.

\bibitem{ranjan2019investigation}
K.~Ranjan, S.~Arulkumaran, G.~Ng, and A.~Sandupatla, ``Investigation of
  self-heating effect on {DC} and {RF} performances in {AlGaN/{GaN HEMTs}} on
  {CVD-diamond},'' \emph{IEEE Journal of the Electron Devices Society}, vol.~7,
  pp. 1264--1269, 2019.

\bibitem{cho2015near}
J.~Cho, Z.~Li, M.~Asheghi, and K.~E. Goodson, ``Near-junction thermal
  management: Thermal conduction in gallium nitride composite substrates,''
  \emph{Annual Review of Heat Transfer}, vol.~18, 2015.

\bibitem{choi2021perspective}
S.~Choi, S.~Graham, S.~Chowdhury, E.~R. Heller, M.~J. Tadjer, G.~Moreno, and
  S.~Narumanchi, ``A perspective on the electro-thermal co-design of ultra-wide
  bandgap lateral devices,'' \emph{Applied Physics Letters}, vol. 119, no.~17,
  p. 170501, 2021.

\bibitem{chatterjee2021electro}
B.~Chatterjee, D.~Ji, A.~Agarwal, S.~H. Chan, S.~Chowdhury, and S.~Choi,
  ``Electro-thermal investigation of {GaN} vertical trench {MOSFETs},''
  \emph{IEEE Electron Device Letters}, vol.~42, no.~5, pp. 723--726, 2021.

\bibitem{kim2023modeling}
T.~Kim, C.~Song, S.~I. Park, S.~H. Lee, B.~J. Lee, and J.~Cho, ``Modeling and
  analyzing near-junction thermal transport in high-heat-flux gan devices
  heterogeneously integrated with diamond,'' \emph{International Communications
  in Heat and Mass Transfer}, vol. 143, p. 106682, 2023.

\bibitem{song2020effect}
C.~Song, J.~Kim, and J.~Cho, ``The effect of {GaN} epilayer thickness on the
  near-junction thermal resistance of {GaN}-on-diamond devices,''
  \emph{International Journal of Heat and Mass Transfer}, vol. 158, p. 119992,
  2020.

\bibitem{chen2019self}
X.~Chen, S.~Boumaiza, and L.~Wei, ``Self-heating and equivalent channel
  temperature in short gate length {GaN HEMTs},'' \emph{IEEE Transactions on
  Electron Devices}, vol.~66, no.~9, pp. 3748--3755, 2019.

\bibitem{Gerrer2021}
T.~Gerrer, J.~Pomeroy, F.~Yang, D.~Francis, J.~Carroll, B.~Loran, L.~Witkowski,
  M.~Yarborough, M.~J. Uren, and M.~Kuball, ``Thermal design rules of
  {AlGaN/GaN}-based microwave transistors on diamond,'' \emph{IEEE Transactions
  on Electron Devices}, vol.~68, no.~4, pp. 1530--1536, 2021.

\bibitem{Bagnall2014}
K.~R. Bagnall, Y.~S. Muzychka, and E.~N. Wang, ``Analytical solution for
  temperature rise in complex multilayer structures with discrete heat
  sources,'' \emph{IEEE Transactions on Components, Packaging and Manufacturing
  Technology}, vol.~4, no.~5, pp. 817--830, 2014.

\bibitem{chen2020modeling}
X.~Chen, S.~Boumaiza, and L.~Wei, ``Modeling bias dependence of self-heating in
  {{GaN HEMTs}} using two heat sources,'' \emph{IEEE Transactions on Electron
  Devices}, vol.~67, no.~8, pp. 3082--3087, 2020.

\bibitem{odabacsi2020improved}
O.~Odaba{\c{s}}{\i}, M.~{\"O}. Akar, B.~B{\"u}t{\"u}n, and E.~{\"O}zbay,
  ``Improved {$T_\text{MAX}$} estimation in {GaN HEMTs} using an equivalent hot
  point approximation,'' \emph{IEEE Transactions on Electron Devices}, vol.~67,
  no.~4, pp. 1553--1559, 2020.

\bibitem{hua2019thermal}
Y.-C. Hua, H.-L. Li, and B.-Y. Cao, ``Thermal spreading resistance in
  ballistic-diffusive regime for {GaN HEMTs},'' \emph{IEEE Transactions on
  Electron Devices}, vol.~66, no.~8, pp. 3296--3301, 2019.

\bibitem{shen2022bias}
Y.~Shen, X.-S. Chen, Y.-C. Hua, H.-L. Li, L.~Wei, and B.-Y. Cao, ``Bias
  dependence of non-{Fourier} heat spreading in {GaN HEMTs},'' \emph{IEEE
  Transactions on Electron Devices}, vol.~70, no.~2, pp. 409--417, 2022.

\bibitem{chatterjee2020nanoscale}
B.~Chatterjee, C.~Dundar, T.~E. Beechem, E.~Heller, D.~Kendig, H.~Kim,
  N.~Donmezer, and S.~Choi, ``Nanoscale electro-thermal interactions in
  {AlGaN/GaN} high electron mobility transistors,'' \emph{Journal of Applied
  Physics}, vol. 127, no.~4, p. 044502, 2020.

\bibitem{chen2021non}
G.~Chen, ``Non-fourier phonon heat conduction at the microscale and
  nanoscale,'' \emph{Nature Reviews Physics}, vol.~3, no.~8, pp. 555--569,
  2021.

\bibitem{bao2018review}
H.~Bao, J.~Chen, X.~Gu, and B.~Cao, ``A review of simulation methods in
  micro/nanoscale heat conduction,'' \emph{ES Energy \& Environment}, vol.~1,
  no.~39, pp. 16--55, 2018.

\bibitem{tang2023phonon}
D.-S. Tang and B.-Y. Cao, ``Phonon thermal transport and its tunability in
  {GaN} for near-junction thermal management of electronics: A review,''
  \emph{International Journal of Heat and Mass Transfer}, vol. 200, p. 123497,
  2023.

\bibitem{hao2018electrothermal}
Q.~Hao, H.~Zhao, Y.~Xiao, and M.~B. Kronenfeld, ``Electrothermal studies of
  {GaN}-based high electron mobility transistors with improved thermal
  designs,'' \emph{International Journal of Heat and Mass Transfer}, vol. 116,
  pp. 496--506, 2018.

\bibitem{SHEN2023124284}
Y.~Shen, H.-A. Yang, and B.-Y. Cao, ``Near-junction phonon thermal spreading in
  {GaN HEMTs}: A comparative study of simulation techniques by full-band phonon
  {Monte Carlo} method,'' \emph{International Journal of Heat and Mass
  Transfer}, vol. 211, p. 124284, 2023.

\bibitem{vermeersch2022thermal}
B.~Vermeersch, R.~Rodriguez, A.~Sibaja-Hernandez, A.~Vais, S.~Yadav,
  B.~Parvais, and N.~Collaert, ``Thermal modelling of {GaN} \& {InP} {RF}
  devices with intrinsic account for nanoscale transport effects,'' in
  \emph{2022 International Electron Devices Meeting (IEDM)}.\hskip 1em plus
  0.5em minus 0.4em\relax IEEE, 2022, pp. 15--3.

\bibitem{jatal2014high}
W.~Jatal, U.~Baumann, K.~Tonisch, F.~Schwierz, and J.~Pezoldt, ``High-frequency
  performance of {GaN} high-electron mobility transistors on {3C-SiC/Si}
  substrates with {Au}-free ohmic contacts,'' \emph{IEEE Electron Device
  Letters}, vol.~36, no.~2, pp. 123--125, 2014.

\bibitem{masetti1983modeling}
G.~Masetti, M.~Severi, and S.~Solmi, ``Modeling of carrier mobility against
  carrier concentration in arsenic-, phosphorus-, and boron-doped silicon,''
  \emph{IEEE Transactions on Electron Devices}, vol.~30, no.~7, pp. 764--769,
  1983.

\bibitem{canali1975electron}
C.~Canali, G.~Majni, R.~Minder, and G.~Ottaviani, ``Electron and hole drift
  velocity measurements in silicon and their empirical relation to electric
  field and temperature,'' \emph{IEEE Transactions on Electron Devices},
  vol.~22, no.~11, pp. 1045--1047, 1975.

\bibitem{hao2017hybrid}
Q.~Hao, H.~Zhao, and Y.~Xiao, ``A hybrid simulation technique for
  electrothermal studies of two-dimensional {GaN-on-SiC} high electron mobility
  transistors,'' \emph{Journal of Applied Physics}, vol. 121, no.~20, p.
  204501, 2017.

\bibitem{warzoha2021applications}
R.~J. Warzoha, A.~A. Wilson, B.~F. Donovan, N.~Donmezer, A.~Giri, P.~E.
  Hopkins, S.~Choi, D.~Pahinkar, J.~Shi, S.~Graham \emph{et~al.},
  ``Applications and impacts of nanoscale thermal transport in electronics
  packaging,'' \emph{Journal of Electronic Packaging}, vol. 143, no.~2, 2021.

\end{thebibliography}
